\documentclass[showpacs,showkeys,preprintnumbers,amssymb,amsmath,aps,superscriptaddress,prb,twocolumn]{revtex4}
\usepackage{graphicx}
\usepackage{dcolumn}
\usepackage{bm}
\usepackage{epsfig}
\usepackage{color}
\begin{document}

\title{Fractional magnetization plateaux of the spin-1/2 Heisenberg orthogonal-dimer chain revisited: 
strong-coupling approach developed from the exactly solved Ising-Heisenberg model}

\author{Taras Verkholyak}
\affiliation{Institute for Condensed Matter Physics, NASU, 1 Svientsitskii Street, L'viv-11, 79011, Ukraine}
\author{Jozef Stre\v{c}ka}
\affiliation{Institute of Physics, Faculty of Science, P. J. \v{S}af\'{a}rik University, Park Angelinum 9, 040 01, Ko\v{s}ice, Slovakia}

\date{\today}

\begin{abstract}
The spin-1/2 Heisenberg orthogonal-dimer chain is considered within the perturbative strong-coupling approach, which is developed from the exactly solved spin-1/2 Ising-Heisenberg orthogonal-dimer chain with the Heisenberg intradimer and the Ising interdimer couplings. Although the spin-1/2 Ising-Heisenberg orthogonal-dimer chain exhibits just intermediate plateaux at zero, one-quarter and one-half of the saturation magnetization, the perturbative treatment up to second order stemming from this exactly solvable model additionally corroborates the fractional one-third plateau as well as the gapless Luttinger spin-liquid phase. It is evidenced that the approximate results obtained from the strong-coupling approach are in an excellent agreement with the state-of-the-art numerical data obtained for the spin-1/2 Heisenberg orthogonal-dimer chain within the exact diagonalization and density-matrix renormalization group method. The nature of individual quantum ground states is comprehensively studied within the developed perturbation theory. 
\end{abstract}


\maketitle

\section{Introduction} 
Fractional magnetization plateaux in low-dimensional quantum Heisenberg spin systems are one of the most fascinating and most targeted topics in the modern condensed matter physics, because they often resemble intriguing quantum ground states with extremely subtle spin order \cite{rich04,hone04,lacr11}. From the experimental point of view, the fractional plateaux have been detected in magnetization curves of a variety of insulating magnetic materials, which mostly provide real-world representatives of zero-dimensional Heisenberg spin clusters \cite{0d-dimer1,0d-dimer2,0d-dimer3,0d-dimer4,0d-dimer5,0d-dimer6,0d-trimer}, one-dimensional Heisenberg spin chains \cite{1d-dc1,1d-dc2,1d-dc3,1d-trimer1,1d-trimer2,1d-trimer3,1d-mixed,1d-tetramer1,1d-tetramer2,1d-tetramer3,1d-s12bac,1d-s1bac,1d-s1ladder1,1d-s1ladder2,sierra99,langari00,langari01} or two-dimensional Heisenberg spin lattices \cite{2d-ssl1,2d-ssl2,2d-ssl3,2d-ssl4,2d-ssl5,2d-triang1,2d-triang2,2d-kagome1,2d-kagome2}.

The fractional magnetization plateaux of one-dimensional quantum Heisenberg chains should satisfy the quantization condition $p (S_u - m_u) \in \mathbb{Z}$ ($p$ is a period of the ground state, $S_u$ and $m_u$ are the total spin and total magnetization per elementary unit, $\mathbb{Z}$ is a set of the integer numbers), which has been derived by Oshikawa, Yamanaka, Affleck (OYA) by extending the Lieb-Schultz-Mattis theorem \cite{lsm,lmt,oya}. It is worthwhile to remark that the OYA criterion provides for a given period of the ground state $p$ a necessary (but not a sufficient) condition for a presence of fractional magnetization plateaux. To the best of our knowledge, all intermediate plateaux of the quantum Heisenberg chains observed to date experimentally are in agreement with the OYA rule when assuming either simple period $p=1$ or just the period doubling $p=2$. For instance, the experimental representatives of the spin-1/2 Heisenberg diamond chain \cite{1d-dc1,1d-dc2,1d-dc3}, the trimerized spin-1/2 Heisenberg chain \cite{1d-trimer1,1d-trimer2,1d-trimer3} and the mixed spin-(1/2,1) Heisenberg chain \cite{1d-mixed} display one-third plateau, the experimental realizations of the tetramerized spin-1/2 Heisenberg chain \cite{1d-tetramer1,1d-tetramer2,1d-tetramer3}, the spin-1/2 Heisenberg bond alternating chain \cite{1d-s12bac} as well as the spin-1 Heisenberg bond alternating chain \cite{1d-s1bac} exhibit one-half plateau, the experimental realization of the spin-1 Heisenberg ladder \cite{1d-s1ladder1,1d-s1ladder2} shows one-quarter plateau, etc. 

From this perspective, it is quite curious that the spin-1/2 Heisenberg orthogonal-dimer (or equivalently dimer-plaquette) chain seems at first sight to contradict the OYA rule, which predicts just its three most pronounced fractional plateaux at zero, one-quarter and one-half of the saturation magnetization when the period of ground state does not exceed doubling of unit cell (i.e. $p=2$). Contrary to this, it has been argued by Schulenburg and Richter on the basis of exact numerical diagonalization data \cite{schulenburg02,sch02} that the spin-1/2 Heisenberg orthogonal-dimer chain exhibits in between one-quarter and one-half plateaux an infinite series of smaller fractional plateaux at $n/(2n+2) = 1/4, 1/3, 3/8, \cdots, 1/2$ of the saturation magnetization corresponding to the ground state with the period $(n+1)$ of unit cell. 
It could be thus concluded that the overall magnetization curve of the spin-1/2 Heisenberg orthogonal-dimer chain is not consistent with any finite period of the ground state.

In this regard, it appears worthwhile to revisit the zero-temperature magnetization curve of the spin-1/2 Heisenberg orthogonal-dimer chain by some another rigorous method, which may capture a formation of the fractional magnetization plateaux of quantum origin. To this end, we will develop in the present work a strong-coupling approach starting from the exactly solved spin-1/2 Ising-Heisenberg orthogonal-dimer chain with the Heisenberg intradimer and Ising interdimer interactions \cite{verkholyak13,v14,pso}. It will be demonstrated that the developed strong-coupling approach actually brings insight into character of individual quantum ground states realized at particular fractional magnetization plateaux. The validity of the method will be also examined by the comparison with the results of the combined numerical approach described in Appendix~\ref{app-num}.

It should be also mentioned that the strong-coupling approach and its modification, the localized-magnon approach, has been recently applied to the asymmetric orthogonal-dimer chain \cite{derzhko13}. However, this study was merely restricted to high magnetic fields and the effect of the asymmetry.

The organization of this paper is as follows. In Sec.~\ref{sec-model} we will introduce the model and suggest its approximate perturbative treatment. The exact solution for the spin-1/2 Ising-Heisenberg orthogonal-dimer chain is formulated within the projection operator technique in Sec.~\ref{sec-IH-model}. In Sec.~\ref{sec-SC} we will develop the strong-coupling approach for the spin-1/2 Heisenberg orthogonal-dimer chain from the exactly solved Ising-Heisenberg model. The main results are summarized in Sec.~\ref{sec-conclusions}.

\section{Heisenberg orthogonal-dimer chain and perturbation method} 
\label{sec-model}

\begin{figure}[b]
\begin{center}
\epsfig{file=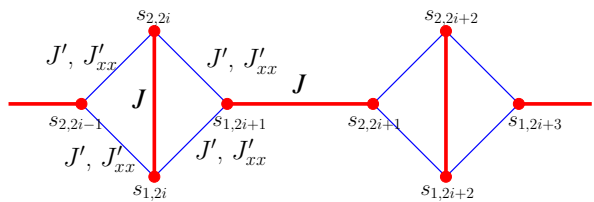, width=0.9\columnwidth}
\end{center}
\caption{(Color online) A schematic representation of the orthogonal-dimer chain along with labeling of intradimer and interdimer coupling constants.}
\label{fig1}
\end{figure}

Let us consider  the spin-1/2 quantum Heisenberg orthogonal-dimer chain given by the Hamiltonian:
\begin{eqnarray}
H&=&\sum_{i=1}^{N}[J({\mathbf s}_{1,i}\cdot{\mathbf s}_{2,i}) - h(s_{1,i}^z+s_{2,i}^z)]
\nonumber\\
&&+\sum_{i=1}^{N/2} J'({\mathbf s}_{1,2i} + {\mathbf s}_{2,2i})\cdot({\mathbf s}_{2,2i-1} + {\mathbf s}_{1,2i+1}),
\label{ham}
\end{eqnarray}
which involves the coupling constants $J$ and $J'$ accounting for the Heisenberg intradimer and interdimer interactions, respectively, in addition to the usual Zeeman's term $h$ (see Fig. \ref{fig1} for a schematic illustration of the considered magnetic lattice). It has been found that the model defined through the Hamiltonian (\ref{ham}) exhibits a singlet-dimer ground state for $J'<0.819J$ \cite{koga00} at zero magnetic field and reveals the peculiar infinite series of the fractional magnetization plateaux in between 1/4 and 1/2 of the saturation magnetization \cite{schulenburg02,sch02}. 

Recently, we have exactly solved the simplified version of this frustrated quantum spin model, the so-called spin-1/2 Ising-Heisenberg orthogonal-dimer chain defined by the Hamiltonian:
\begin{eqnarray}
H&=&\sum_{i=1}^{N}[J({\mathbf s}_{1,i}\cdot{\mathbf s}_{2,i}) - h(s_{1,i}^z+s_{2,i}^z)]
\nonumber\\
&&+\sum_{i=1}^{N/2} J'({s}_{1,2i}^z + {s}_{2,2i}^z)({s}_{2,2i-1}^z + {s}_{1,2i+1}^z),
\label{hamih}
\end{eqnarray}
which takes into account the Heisenberg intradimer interaction $J$ and the Ising interdimer interaction $J'$  \cite{verkholyak13,v14}. The only difference between two models lies in replacing the Heisenberg interdimer coupling in the Hamiltonian (\ref{ham}) through the Ising interdimer coupling in the Hamiltonian (\ref{hamih}). The simplified spin-1/2 Ising-Heisenberg orthogonal-dimer chain (\ref{hamih}) can be rigorously solved either by the transfer-matrix method \cite{verkholyak13,v14} or the mapping transformation technique \cite{pso}, whereas this model still exhibits some common features with its full Heisenberg counterpart like intermediate magnetization plateaux at one-quarter and one-half of the saturation magnetization. However, the exactly solved Ising-Heisenberg model given by the Hamiltonian (\ref{hamih}) does not reproduce neither an infinite series of the fractional magnetization plateaux in between one-quarter and one-half of the saturation magnetization nor an existence of the Tomonaga-Luttinger spin-liquid phase above the intermediate one-half plateau. Instead it shows the macroscopically degenerate ground-state manifold at each critical field accompanied with the magnetization jump \cite{v14}. This fact enables us to develop an approximate theory for the spin-1/2 Heisenberg orthogonal-dimer chain based on the exactly solved spin-1/2 Ising-Heisenberg orthogonal-dimer chain when treating the $XY$ part of the interdimer coupling perturbatively. 

To this end, let us decompose the total Hamiltonian (\ref{ham}) of the spin-1/2 Heisenberg orthogonal-dimer chain into two parts
\begin{eqnarray}
\label{gen_ham1}
H&{=}& H^{(0)} + H^{(1)} = \sum_{i=1}^N H^{(0)}_i +  \sum_{i=1}^N H^{(1)}_i, 
\end{eqnarray}
where the former unperturbed (ideal) part $H^{(0)}$ corresponds to the exactly solved spin-1/2 Ising-Heisenberg orthogonal-dimer chain \cite{verkholyak13} rewritten as
\begin{eqnarray}
H^{(0)}_{2i}&{=}&
J({\mathbf s}_{1,2i}\cdot{\mathbf s}_{2,2i}) - h_c(s_{1,2i}^z+s_{2,2i}^z),
\nonumber\\
H^{(0)}_{2i+1}&{=}&
J'[(s_{1,2i}^z{+}s_{2,2i}^z)s_{1,2i{+}1}^z{+}s_{2,2i{+}1}^z(s_{1,2i{+}2}^z{+}s_{2,2i{+}2}^z)]
\nonumber\\&&
{+}J({\mathbf s}_{1,2i{+}1}\cdot{\mathbf s}_{2,2i{+}1})
{-}h_c(s_{1,2i{+}1}^z{+}s_{2,2i{+}1}^z),
\label{unper}
\end{eqnarray}
while the latter perturbed part $H^{(1)}$ contains all remaining terms from the total Hamiltonian (\ref{ham}) 
of the spin-1/2 Heisenberg orthogonal-dimer chain
\begin{eqnarray}
H^{(1)}_{2i}&{=}& (h_c - h)(s_{1,2i}^z+s_{2,2i}^z)
\nonumber\\
&&+J'_{xx}\sum_{\alpha=x,y}(s_{1,2i}^\alpha + s_{2,2i}^\alpha)(s_{2,2i-1}^\alpha + s_{1,2i+1}^\alpha),
\nonumber\\
H^{(1)}_{2i+1}&{=}& (h_c - h)(s_{1,2i+1}^z+s_{2,2i+1}^z).
\end{eqnarray}
It is noteworthy that the perturbed Hamiltonian $H^{(1)}$ includes except the $XY$ part of the interdimer coupling also difference between the true magnetic field $h$ and its respective critical value $h_c$, around each of which one should separately perform the perturbative expansion due to a macroscopic degeneracy of the ground-state manifold of the spin-1/2 Ising-Heisenberg orthogonal-dimer chain \cite{verkholyak13,v14}. 
The macroscopic degeneracies at the critical fields and their values will be given and discussed in the next section.
Though we have singled out the $XY$-part of the interdimer interaction $J'_{xx}$ explicitly, the isotropic limit of the quantum Heisenberg model will be later recovered by putting $J'_{xx}=J'$ in all final expressions. Besides, our further consideration will be limited only to the most interesting case with the antiferromagnetic interactions $J,J'>0$ under the simultaneous constraint $J'<0.819J$, which favors the singlet-dimer phase as the zero-field ground state of the spin-1/2 Heisenberg orthogonal-dimer chain \cite{koga00}.

\section{Exact solution of the Ising-Heisenberg orthogonal-dimer chain in terms of the projection operators}
\label{sec-IH-model}

Although the exact solution of the spin-1/2 Ising-Heisenberg orthogonal-dimer chain given by the Hamiltonian (\ref{hamih}) [or equivalently by the Hamiltonians (\ref{unper})] have been already reported by two independent methods, i.e. the transfer-matrix method \cite{verkholyak13,v14} and the mapping transformation technique \cite{pso}, it appears worthwhile to rederive it by making use of the projection operators in view of a subsequent development of the perturbative strong-coupling approach. For this purpose, let us introduce first the dimer-state basis
\begin{eqnarray}
&&|0\rangle_i=\frac{1}{\sqrt2}(|\!\uparrow\rangle_{1,i}|\!\downarrow\rangle_{2,i}-|\!\downarrow\rangle_{1,i}|\!\uparrow\rangle_{2,i}),
\nonumber\\
&&|1\rangle_i=|\!\uparrow\rangle_{1,i}|\!\uparrow\rangle_{2,i},
\nonumber\\
&&|2\rangle_i=\frac{1}{\sqrt2}(|\!\uparrow\rangle_{1,i}|\!\downarrow\rangle_{2,i}+|\!\downarrow\rangle_{1,i}|\!\uparrow\rangle_{2,i}),
\nonumber\\
&&|3\rangle_i=|\!\downarrow\rangle_{1,i}|\!\downarrow\rangle_{2,i},
\label{dimerbasis}
\end{eqnarray}
and the corresponding projection operators \cite{parkinson79,Xian94,Xian95}
\begin{eqnarray}
A^{ab}_i=|a\rangle_i\langle b|_i.
\label{proj-operators}
\end{eqnarray}
One can find the representation of spin operators through the introduced projection operators $A^{ab}_i$ (the explicit correspondence is given in Appendix \ref{app-projection}) and rewrite  in terms of these operators the local Hamiltonians (\ref{unper}) pertinent to the vertical and horizontal Heisenberg dimers  (see Fig.~\ref{fig1}):
\begin{eqnarray}
H^{(0)}_{2i}&=&J(\frac{1}{4}-A_{2i}^{00}) - h_c(A_{2i}^{11}-A_{2i}^{33}),
\label{H0_even} 
\\
H^{(0)}_{2i+1}&=&J(\frac{1}{4}-A_{2i+1}^{00}) - h_c(A_{2i+1}^{11}-A_{2i+1}^{33})
\nonumber\\&&
+\frac{J'}{2}\left[(S_{2i}^z+S_{2i+2}^z)(A_{2i+1}^{11}-A_{2i+1}^{33}) 
\right.\nonumber\\&&\left.
+ (S_{2i}^z-S_{2i+2}^z)(A_{2i+1}^{20}+A_{2i+1}^{02}) \right]. 
\label{H0_odd}
\end{eqnarray}
Here, $S_{2i}^z=s_{1,2i}^z+s_{2,2i}^z=A_{2i}^{11}-A_{2i}^{33}$ denotes the $z$-component of the total spin on $2i$th vertical dimer, whereas an explicit form of the total spin $S_{2i}^z$ and $S_{2i+2}^z$ on two neighboring vertical dimers has been retained in Eq. (\ref{H0_odd}) for the sake of compactness. It is quite evident that the Hamiltonians $H^{(0)}_{2i}$ of the vertical dimers (\ref{H0_even}) are already diagonal in the dimer representation, while the Hamiltonians $H^{(0)}_{2i+1}$ of the horizontal dimers (\ref{H0_odd}) can be diagonalized by a unitary transformation:
\begin{eqnarray}
&&U_{2i+1}=(A_{2i+1}^{11}+A_{2i+1}^{33}) + \cos\frac{\alpha_{2i+1}}{2}(A_{2i+1}^{00}+A_{2i+1}^{22})
\nonumber\\
&&+ \sin\frac{\alpha_{2i+1}}{2}(A_{2i+1}^{20}-A_{2i+1}^{02}),
\nonumber\\
&&\cos\alpha_{2i+1}=\frac{J}{\sqrt{J^2+J'^2(S_{2i}^z - S_{2i+2}^z)^2} }, \;\,\,\,
\nonumber\\
&&\sin\alpha_{2i+1}=\frac{J'(S_{2i}^z - S_{2i+2}^z)}{\sqrt{J^2+J'^2(S_{2i}^z - S_{2i+2}^z)^2}}. 
\label{u-transform}
\end{eqnarray}
It should be stressed that $\cos\alpha_{2i+1}$ and $\sin\alpha_{2i+1}$ depend on eigenvalues of the operators $S_{2i}^z$, $S_{2i+2}^z$, and they can be reduced to an algebraic form using the van der Waerden identity (see e.g. Refs.~\onlinecite{tucker94,strecka15}).
The explicit expressions for $\cos\frac{\alpha_{2i+1}}{2}$ and $\sin\frac{\alpha_{2i+1}}{2}$ is given in Appendix \ref{app-UT}. Apparently, two polarized triplet states $|1\rangle$ and $|3\rangle$ are invariant under the unitary transformation (\ref{u-transform}), while the singlet $|0\rangle$ and the zero-component of the triplet state $|2\rangle$ are mutually entangled to a more complex quantum state:
\begin{eqnarray}
|\tilde 0\rangle_{2i+1}&=&U_{2i+1}|0\rangle_{2i+1}
\nonumber\\
&=&\cos\frac{\alpha_{2i+1}}{2}|0\rangle_{2i+1} + \sin\frac{\alpha_{2i+1}}{2}|2\rangle_{2i+1},
\nonumber\\
|\tilde 2\rangle_{2i+1}&=&U_{2i+1}|2\rangle_{2i+1}
\nonumber\\
&=&{-}\sin\frac{\alpha_{2i+1}}{2}|0\rangle_{2i+1} {+} \cos\frac{\alpha_{2i+1}}{2}|2\rangle_{2i+1}.
\end{eqnarray}
After performing the local unitary transformation (\ref{u-transform}) one consequently obtains the diagonal form of the Hamiltonian 
$\tilde{H}^{(0)}_{2i+1}=U_{2i+1}H^{(0)}_{2i+1}U_{2i+1}^+$ 
of the ($2i+1$)st horizontal dimer
\begin{widetext}
\begin{eqnarray}
&&\tilde{H}^{(0)}_{2i+1}=
\left(\frac{J'(S_{2i}^z + S_{2i+2}^z)}{2} -h_c \right)(A_{2i+1}^{11}-A_{2i+1}^{33}) 
+J(\frac{1}{4}-A_{2i+1}^{00}) + \frac{1}{2}I(|S_{2i}^z - S_{2i+2}^z|)(A_{2i+1}^{22}-A_{2i+1}^{00}),
\nonumber\\
&&I(|S_{2i}^z - S_{2i+2}^z|)=\delta(|S_{2i}^z - S_{2i+2}^z|-1)(\sqrt{J^2+J'^2}-J)
+\delta(|S_{2i}^z - S_{2i+2}^z|-2)(\sqrt{J^2+4J'^2}-J).
\label{hamdh}
\end{eqnarray}
\end{widetext}
Here $\delta(\dots)$ is a symbolic notation of Kronecker delta. Its algebraic representation through the spin and projection operators can be found in Appendix~\ref{app-UT} (see Eq.~(\ref{kron-delta})).

Using this procedure, the total Hamiltonian (\ref{hamih}) of the spin-1/2 Ising-Heisenberg orthogonal-dimer chain has been put into a fully diagonal form and the ground state of the model can be easily found by minimizing a sum of its local diagonal parts (\ref{H0_even}) and (\ref{hamdh}) (see also Ref.~\onlinecite{verkholyak13}). By inspection, one finds just four different ground states in the investigated parameter space $J'<0.819J$ and $h>0$, namely,
\begin{itemize}
\item singlet-dimer (SD) phase: $|{\rm SD}\rangle= \displaystyle\prod_{i=1}^N |\tilde 0\rangle_i$,
\item modulated ferrimagnetic (MFI) phase:\\ 
$|{\rm MFI}\rangle=\displaystyle\prod_{i=1}^{N/4}
\left\{
\begin{array}{l}
|\tilde 0\rangle_{4i-3}|\tilde 0\rangle_{4i-2}|\tilde 0\rangle_{4i-1}|\tilde 1\rangle_{4i},\\
|\tilde 0\rangle_{4i-3}|\tilde 1\rangle_{4i-2}|\tilde 0\rangle_{4i-1}|\tilde 0\rangle_{4i},
\end{array} 
\right.$\\
\item staggered bond (SB) phase:\\ 
$|{\rm SB}\rangle=\displaystyle\prod_{i=1}^{N/2} 
\left\{
\begin{array}{l}
|\tilde 1\rangle_{2i-1}|\tilde 0\rangle_{2i},\\
|\tilde 0\rangle_{2i-1}|\tilde 1\rangle_{2i},\end{array} 
\right.$,
\item saturated (SAT) phase: $|{\rm SAT}\rangle=\displaystyle\prod_{i=1}^N |\tilde 1\rangle_i$.
\end{itemize}
It is worthwhile to recall that the ground state is macroscopically degenerate at the critical fields, where the magnetization discontinuously jumps due to successive field-induced (first-order) phase transitions SD$\to$MFI$\to$SB$\to$SAT upon strengthening of the magnetic field. The explicit form of the critical fields corresponding to the relevant ground-state phase boundaries were found in Ref. \onlinecite{verkholyak13}:
\begin{itemize}
\item SD-MFI:
$h_{c1}=2J-\sqrt{J^2+J'^2}$,
\item MFI-SB:
$h_{c2}=\sqrt{J^2+J'^2}$, 
\item SB-SAT:
$h_{c3}=J+J'$.
\end{itemize}
The ground-state manifold along with its macroscopic degeneracy at a given critical field can be obtained from the condition of the phase coexistence 
of both individual ground states. For instance, all horizontal dimers have to be in the singlet-like state $|\tilde 0\rangle_{2i-1}$ at SD-MFI boundary, while the polarized triplet states $|\tilde 1\rangle_{2i}$ can be randomly distributed on the vertical dimers on assumption that the hard-core repulsion between the nearest-neighboring polarized states on the vertical dimers is fulfilled (the remaining vertical dimers have to be in the singlet-like state $|\tilde 0\rangle_{2i}$). Thus, the ground-state manifold at SD-MFI phase boundary can be defined through the following projection operator:
\begin{eqnarray}
\label{b_SD-MFI}
P_{SD-MFI}=\prod_{i=1}^{N/2} A_{2i-1}^{00}(A_{2i}^{00}+A_{2i-2}^{00}A_{2i}^{11}A_{2i+2}^{00}). 
\end{eqnarray}
Similarly, the ground-state manifold at SB-SAT boundary can be built from any random configuration of the singlet-like states $|\tilde 0\rangle_{2i-1}$ and $|\tilde 0\rangle_{2i}$ on the horizontal and vertical dimers, which satisfies the hard-core repulsion between the singlet-like states on the nearest-neighbor dimers (the remaining dimers should occupy the polarized triplet states $|\tilde 1\rangle_{2i-1}$ and $|\tilde 1\rangle_{2i}$). 
The ground-state manifold at SB-SAT phase boundary is thus given by the following projection operator:
\begin{eqnarray}
\label{b_SB-SAT}
P_{SB-SAT}=\prod_{i=1}^{N} (A_{i}^{11}+A_{i-1}^{11}A_{i}^{00}A_{i+1}^{11}). 
\end{eqnarray}
The situation at MFI-SB phase boundary is much more intricate and it does not allow such a transparent representation. However, the ground-state manifold at MFI-SB phase boundary can be defined through the projection operator as follows:
\begin{eqnarray}
\label{b_MFI-SB}
&&P_{MFI-SB}=\prod_{i=1}^{N/2} [A_{2i-2}^{00}A_{2i-1}^{11}A_{2i}^{00}
\nonumber\\
&&+A_{2i-1}^{00}(A_{2i-2}^{11}A_{2i}^{00} + A_{2i-2}^{00}A_{2i}^{11} + A_{2i-2}^{11}A_{2i}^{11} )].
\end{eqnarray}

\section{Strong-coupling approach developed from the exactly solved Ising-Heisenberg model}
\label{sec-SC}

The strong-coupling approach is based on the many-body perturbation theory (see e.g. Ref.~\onlinecite{b_fulde91,b_mila11}), where the unperturbed $H^{(0)}$ and perturbed $H^{(1)}$ parts of the Hamiltonian can be singled out:
\begin{eqnarray}
H=H^{(0)} + \lambda H^{(1)}
\end{eqnarray}
and the eigenvalue problem for the ideal part $H^{(0)}|\Phi_i\rangle=E_i^{(0)}|\Phi_i\rangle$ becomes exactly tractable. If $P$ is the projection operator on a ground state $|\Phi_0\rangle$ of the unperturbed model subspace $H^{(0)}$ and $Q=1-P$, the perturbative expansion can be formally 
found out for the effective Hamiltonian acting in the projected subspace $P$:
\begin{eqnarray}
H_{eff}&=&PHP 
\nonumber\\
&+& \lambda^2PH^{(1)}R_s\sum_{n=0}^{\infty}
\left( (QH^{(1)}{-}\delta E_0)R_s\right)^nQH^{(1)}P, 
\nonumber\\
 R_s&=&Q\frac{1}{E_0^{(0)}-H^{(0)}}=\sum_{m\neq 0}\frac{|\Phi_m\rangle\langle\Phi_m|}{E_0^{(0)}-E_m^{(0)}}
\nonumber\\
&=&\sum_{m\neq 0}\frac{Q_m}{E_0^{(0)}-E_m^{(0)}},
\label{eq_pert}
\end{eqnarray}
where $\delta E_0=E_0-E_0^{(0)}$. Note that the perturbative expansion (\ref{eq_pert}) is still exact, but one usually has to truncate it due to computational difficulties arising out from higher-order contributions $H_{eff}^{(n)}$ of the effective Hamiltonians. In the present work we will restrict ourselves to the second-order perturbative expansion, which will take into account the zeroth-, first- and the second-order contributions to the effective Hamiltonian: $H_{eff}^{(0)}=PH^{(0)}P$, $H_{eff}^{(1)}=\lambda PH^{(1)}P$ and $H_{eff}^{(2)}=\lambda^2 PH^{(1)} R_s H^{(1)}P$, respectively.
In what follows we will develop the perturbation theory for the spin-1/2 Heisenberg orthogonal-dimer chain from the exactly solved spin-1/2 Ising-Heisenberg orthogonal-dimer chain by considering separately the macroscopically degenerate ground-state manifold at each its phase boundary.

\subsection{SD-MFI boundary}
The phase boundary between SD and MFI ground states of the spin-1/2 Ising-Heisenberg orthogonal-dimer chain is defined by the critical field $h_{c1}$ and the projection operator to the macroscopically degenerate ground-state manifold is given by Eq.~(\ref{b_SD-MFI}). The straightforward application of Eq.~(\ref{eq_pert}) results in the first-order term:
\begin{eqnarray}
H_{eff}^{(1)}&=&P_{SD-MFI}\tilde{H}^{(1)}P_{SD-MFI}
\nonumber\\
&=&(h_{c1}-h)\left(\sum_{i=1}^{N/2}A_{2i}^{11}\right) P_{SD-MFI},
\end{eqnarray}
where $\tilde{H}^{(1)}=U {H}^{(1)} U$ is the unitary-transformed perturbation operator. The second-order term $H_{eff}^{(2)}$ requires the calculation of the matrix elements and is much more involved (see the details of the calculations in Appendix \ref{app-SD-MFI}):
\begin{eqnarray}
&&\hspace{-25pt}H_{eff}^{(2)}=\sum_{i=1}^{N/2} h^{(2)} A_{2i}^{11}P_{SD-MFI},
\nonumber\\
&&\hspace{-25pt}h^{(2)}{=}{-}\frac{(J'_{xx})^2(c_1^+ {+} c_1^-)^2}{2}\left\{
\frac{(c_1^+)^2}{\sqrt{J^2{+}J'^2}} {+} \frac{(c_1^-)^2}{J{+}\sqrt{J^2{+}J'^2}}
\right\}, \label{socsdmfi}
\\
&&\hspace{-25pt}c_1^{\pm}=\frac{1}{\sqrt2}\sqrt{1\pm\frac{J}{\sqrt{J^2+J'^2} } }.
\end{eqnarray}
Summing up all contributions up to second order we get the following effective Hamiltonian:
\begin{eqnarray}
H_{eff}&=&\sum_{i=1}^{N/2} (h_{c1}+h^{(2)}-h) A_{2i}^{11}P_{SD-MFI}.
\label{efsm1}
\end{eqnarray}
Obviously, the effective Hamiltonian (\ref{efsm1}) is essentially one-dimensional classical model with a simple mapping correspondence to the lattice-gas model, which can be established by considering the singlet-like (polarized triplet) states on the vertical dimers as being empty (filled) sites: $|0\rangle \rightarrow \mbox{empty site}$, $|1\rangle \rightarrow \mbox{filled site}$, $A_{2i}^{00}=(1-n_i)$, $A_{2i}^{11}=n_i$. The effective Hamiltonian in the lattice-gas representation is extraordinarily simple and it satisfies the hard-core constraint for the polarized triplet states 
on the nearest-neighbor vertical dimers as dictated by the projection operator (\ref{b_SD-MFI}):
\begin{eqnarray}
H_{eff}=\sum_{i=1}^{N/2} (h_{c1}+h^{(2)}-h) (1-n_{i-1})n_i(1-n_{i+1}).
\end{eqnarray}
The ground state corresponds either to the lattice-gas model with all empty sites for $h<h_{c1}+h^{(2)}$ or the half-filled case for $h>h_{c1}+h^{(2)}$. The former condition with all empty sites ($n_i=0$ for all $i$) is consistent with SD ground state of the original spin model (\ref{ham}), while the latter condition with a regular alternation of empty and filled sites apparently corresponds to MFI phase. It could be thus concluded that the second-order perturbative expansion around SD-MFI phase boundary does not create any novel ground state, but it only renormalizes the critical field $h_{c1}+h^{(2)}$ of a discontinuous phase transition between SD and MFI ground states accompanied with the magnetization jump from zero to one-quarter of the saturation magnetization. It is quite evident from Eq. (\ref{socsdmfi}) that the second-order correction to the first critical field is negative ($h^{(2)}<0$), which is consequently shifted to lower values of the magnetic field in an excellent accordance with the state-of-the-art numerical data obtained from the density-matrix renormalization group (DMRG) and exact diagonalization (ED) calculations described in Appendix~\ref{app-num} (c.f. Figs. \ref{fig_phasediag} and \ref{fig_mag}).
\begin{figure}
\begin{center}
\epsfig{file=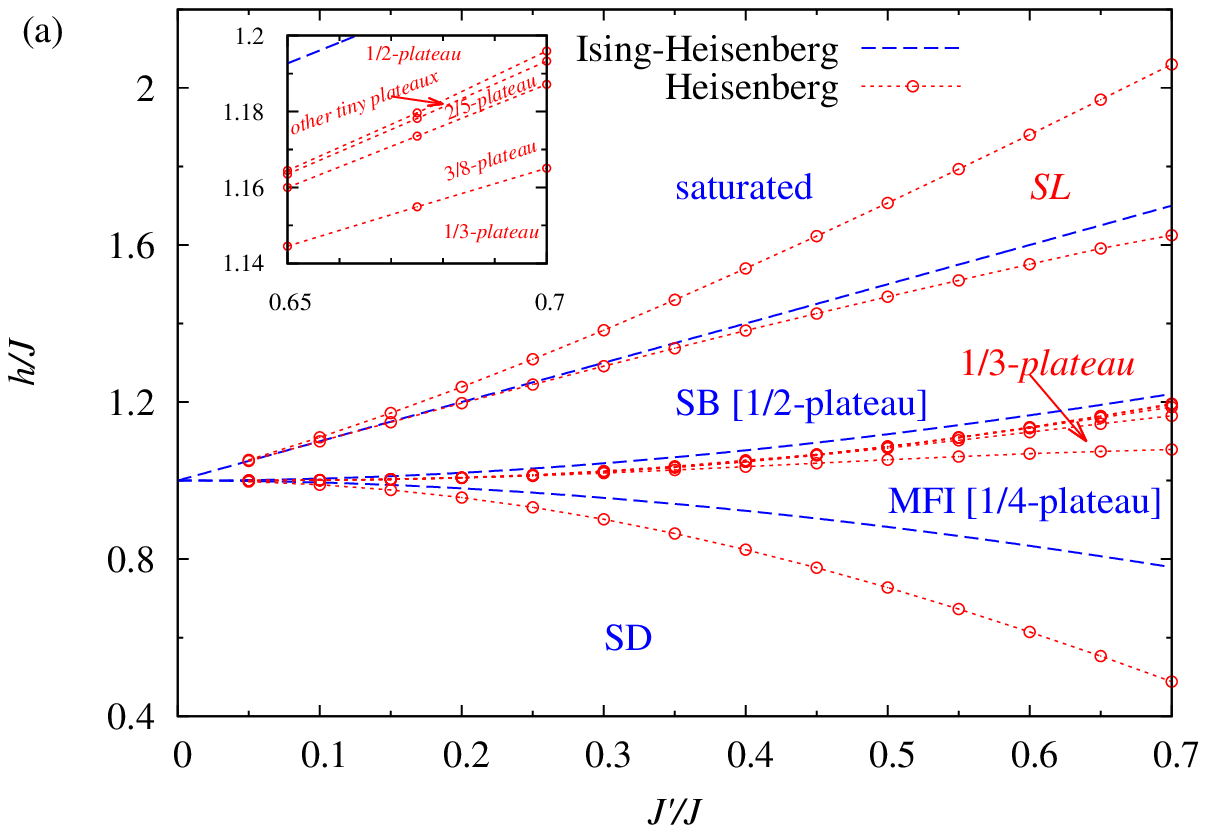, width=1.\columnwidth}
\epsfig{file=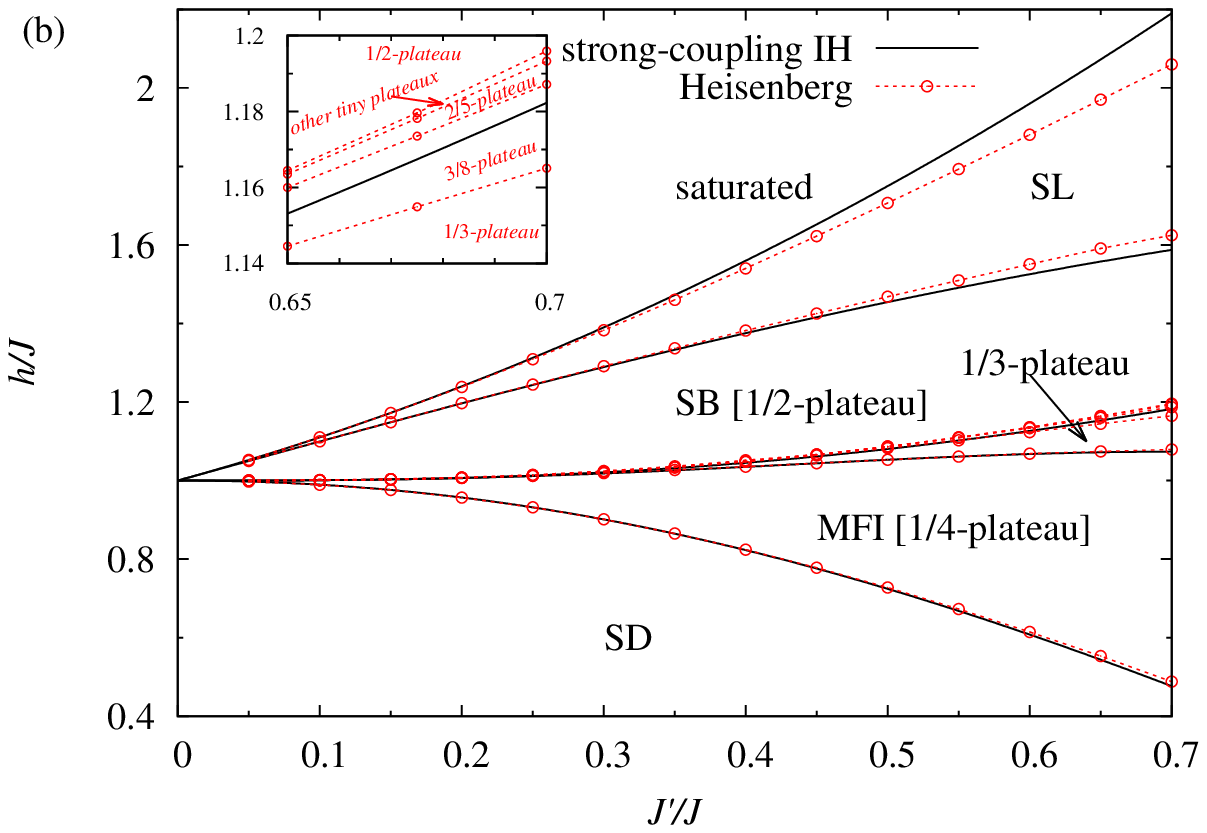, width=1.\columnwidth}
\end{center}
\caption{(Color online) The ground-state phase diagram of the spin-1/2 orthogonal-dimer chain in the $J'/J-h/J$ plane: (a) the exact results for the Ising-Heisenberg model (broken lines) versus the numerical data for the Heisenberg model (dotted lines with symbols); (b) the perturbative strong-coupling approach (solid lines) versus the numerical data (dotted lines with symbols) for the Heisenberg model. The  numerical method is specified in Appendix~\ref{app-num}.}
\label{fig_phasediag}
\end{figure}
\begin{figure}[h]
\begin{center}
\epsfig{file=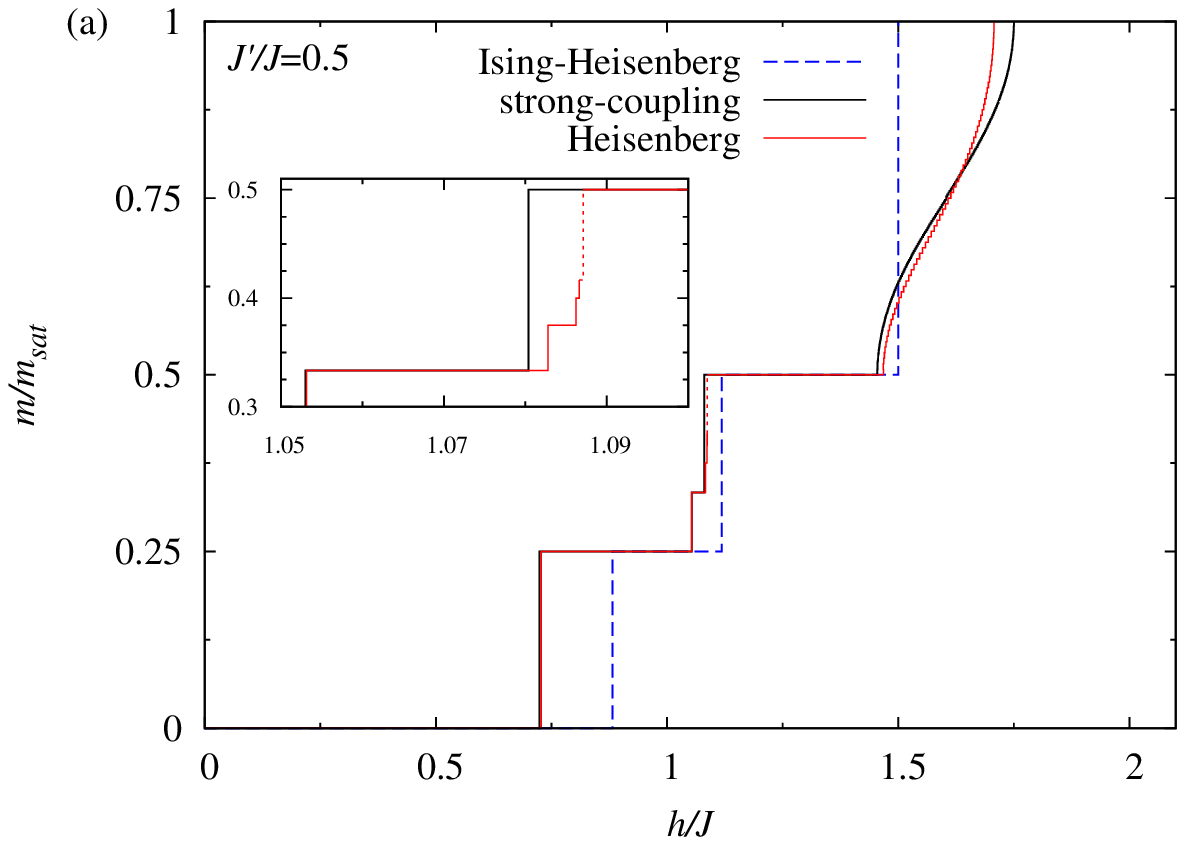, width=1.\columnwidth}
\epsfig{file=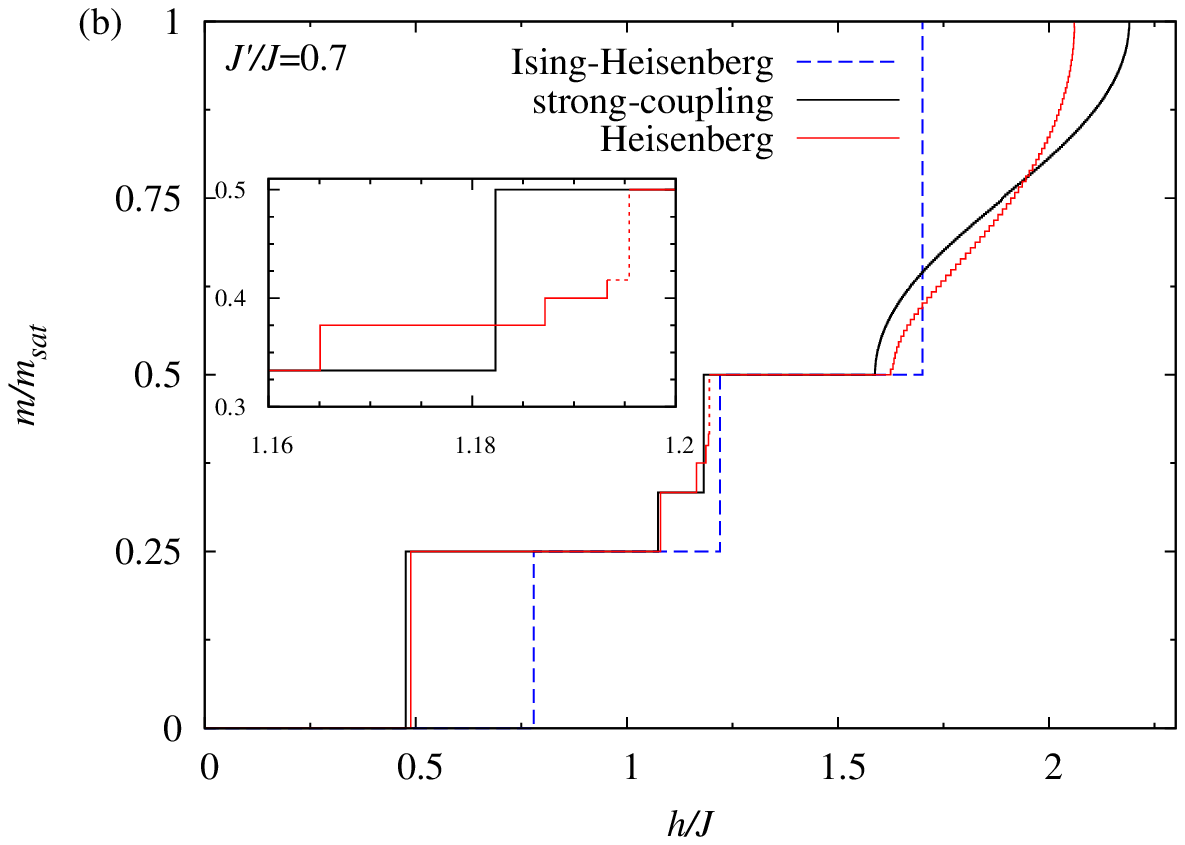, width=1.\columnwidth}
\end{center}
\vspace*{-0.3cm}
\caption{(Color online) A comparison of the exact zero-temperature magnetization curve of the spin-1/2 Ising-Heisenberg orthogonal-dimer chain (broken lines) with the zero-temperature magnetization curve of the spin-1/2 Heisenberg orthogonal-dimer chain obtained within the strong-coupling approach (black solid lines) and the numerical method specified in Appendix~\ref{app-num} (red solid lines) for two different values of the interaction ratio: (a) $J'/J=0.5$; (b) $J'/J=0.7$. In the insets we show in an enlarged scale the field region with an infinite sequence of the fractional plateaux, whereas the dotted line denotes the case when tiny magnetization plateaux at $n/(2n+2)$ become indistinguishable for $n>5$ within the chosen scale.}
\label{fig_mag}
\end{figure}

\subsection{MFI-SB boundary}
The phase boundary between MFI and SB ground states of the spin-1/2 Ising-Heisenberg orthogonal-dimer chain occurs at the second critical field $h_{c2}$, at which the projector (\ref{b_MFI-SB}) determines the macroscopically degenerate ground-state manifold. One may use the same procedure as before in order to get the effective Hamiltonian. The first-order contribution to the effective Hamiltonian is determined by the diagonal elements of the perturbed part of the Hamiltonian:
\begin{eqnarray}
H_{eff}^{(1)}
=(h_{c2}-h)\left(\sum_{i=1}^{N/2}A_{2i}^{11}\right) P_{MFI-SB}.
\end{eqnarray}
After cumbersome calculations one gets of the following result for the second-order contribution to the effective Hamiltonian (see Appendix \ref{app-MFI-SB} for further details):
\begin{eqnarray}
&&H_{eff}^{(2)}= \sum_{i=1}^{N/2}
\left\{
D_{00}A_{2i-2}^{00}A_{2i+2}^{00}
+D_{11}A_{2i-2}^{11}A_{2i+2}^{11}
\right.\nonumber\\&&\left.
+D_{01}(A_{2i-2}^{11}A_{2i+2}^{00}+A_{2i-2}^{00}A_{2i+2}^{11})
\right\}A_{2i}^{11} P_{MFI-SB},
\nonumber\\
&& D_{00}{=}{-}\frac{(J'_{xx})^2(c_1^+ {+} c_1^-)^2}{2}
\left(\frac{(c_1^+)^2}{\sqrt{J^2{+}J'^2}} {+} \frac{(c_1^-)^2}{J{+}\sqrt{J^2{+}J'^2}}
\right),
\nonumber\\
&& D_{11}=-(J'_{xx})^2
\left(\frac{(c_1^+)^2}{3J+J'-\sqrt{J^2+J'^2}} 
\right.\nonumber\\&&\left.
+ \frac{(c_1^-)^2}{3J+J'+\sqrt{J^2+J'^2}}
\right),
\nonumber\\
&& D_{01}=-\frac{(J'_{xx})^2}{4}\left\{
\frac{2(c_1^+)^2}{J+J'+\sqrt{J^2+J'^2}} 
\right.\nonumber\\&&
+ \frac{2(c_1^-)^2}{3J+J'+\sqrt{J^2+J'^2}}
\nonumber\\&&\left.
+(c_1^+ + c_1^-)^2
\left(\frac{(c_1^+)^2}{J} + \frac{(c_1^-)^2}{J+\sqrt{J^2+J'^2}}
\right)
\right\}.
\label{hefmfisb}
\end{eqnarray}
Since all three expansion coefficients are negative ($D_{00},D_{11},D_{01}<0$) one generally has $\langle H_{eff}^{(2)}\rangle\leq 0$ and $\langle H_{eff}^{(2)}\rangle= 0$ if horizontal dimers are in the singlet-like states. 
It is quite straightforward to show that the ground state corresponds to the state with $|0\rangle_{2i-1}$, i.e. $A^{11}_{2i-1}|GS\rangle=0$. Therefore, the states with the polarized horizontal triplets can be excluded from the consideration 
if we are seeking only for the ground state. Let us introduce the notation $A_{2i}^{00}=n_i$, $A_{2i}^{11}=(1-n_i)$ in order to rewrite the Hamiltonian (\ref{hefmfisb}) in the lattice-gas representation:
\begin{eqnarray}
H_{eff}&{=}&\sum_{i=1}^{N/2} \left[
D_{11}{+}h_{c2}-h 
{+}(2D_{01}{-}3D_{11} {-}h_{c2} {+}h)n_i
\right.\nonumber\\&+&\left.
(D_{00}+D_{11}-2D_{01})n_{i-1}n_{i+1}
\right]P_{MFI-SB}.
\label{heflgmfisb}
\end{eqnarray}
Similarly to the previous case one obtains the classical effective Hamiltonian with the hard-core potential, but there also appears some additional next-nearest-neighbor interaction. When looking for the lowest-energy states of the effective lattice-gas model given by the Hamiltonian  (\ref{heflgmfisb}), one finds three different ground states either with empty, one-third-filled or half-filled states upon varying the external magnetic field. These lowest-energy states correspond to the fractional plateaux at the one-half, one-third or one-quarter of the saturation magnetization, 
whereas two conditions of a phase coexistence determine the critical fields associated with the respective magnetization jumps:
\begin{eqnarray}
&&h(1/4\to1/3)=h_{c2}+4D_{01}-3D_{00},
\nonumber\\
&&h(1/3\to1/2)=h_{c2}+3D_{11}-2D_{01}.
\end{eqnarray}
The perturbation expansion around the MFI-SB phase boundary thus surprisingly verifies an existence of the fractional one-third magnetization plateau, which is totally absent in a zero-temperature magnetization curve of the spin-1/2 Ising-Heisenberg orthogonal-dimer chain (see Fig.~\ref{fig_mag}(a)). Besides, the method also brings insight into a microscopic nature of the spin arrangement realized within the 1/3-plateau, in which singlet-like states are spread over all horizontal dimers and each third vertical dimer. It can be seen from Fig.~\ref{fig_phasediag}(b) that the developed perturbation theory predicts the critical field $h(1/4\to1/3)$ between 1/4- and 1/3-plateaux in a perfect agreement with the numerical results (see Appendix~\ref{app-num}), while the other critical field $h(1/3\to1/2)$ between 1/3- and 1/2-plateaux lies in a middle of the tiny region involving an infinite sequence of the fractional magnetization plateaux $n/(2n+2)$. There are strong indications that the other tiny fractional magnetization plateaux could be also recovered if the perturbation expansion would be performed up to higher orders. In this case, the repulsion between further neighbors in the lattice-gas representation appears leading to the one-quarter-filled, one-fifth-filled, \ldots, states. These states correspond to the 3/8-, 2/5-plateaux in the original spin model using the relation $m/m_{sat}=(1-\langle n_i\rangle)/2$.

\subsection{SB-SAT boundary}
\begin{figure}
\begin{center}
\epsfig{file=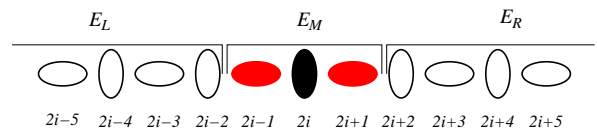, width=0.9\columnwidth}
\end{center}
\caption{(Color online) A schematic representation of the effective quantum lattice-gas model (\ref{eff_SB-SAT}). The black painted ellipse denotes the singlet state of the vertical dimer, the red painted ellipses denote neighboring horizontal dimers where the singlet state is forbidden due to the hard-core repulsion.}
\label{fig_eff-scheme}
\end{figure}

The phase boundary between SB and SAT ground states of the spin-1/2 Ising-Heisenberg orthogonal-dimer chain represents quite exceptional case, because  the perturbative strong-coupling approach will lead in this specific case to the effective Hamiltonian of a quantum nature. The critical field relevant to this phase boundary is given by $h_{c3}=J+J'$, while the macroscopically degenerate ground-state manifold is defined by the projection operator (\ref{b_SB-SAT}). Applying the perturbation theory one obtains the following first-order contribution to the effective Hamiltonian:
\begin{eqnarray}
H_{eff}^{(1)}= (h_{c3}-h)\left(\sum_{i=1}^{N}A_{i}^{11}\right) P_{SB-SAT}.
\end{eqnarray}
After tedious calculations (see Appendix \ref{app-SB-SAT}) one may also find the second-order perturbation term:
\begin{widetext}
\begin{eqnarray}
H_{eff}^{(2)}=\sum_{i=1}^{N/2}P_{SB-SAT}\{
L_1(A_{2i-1}^{00}A_{2i+1}^{11}+A_{2i-1}^{11}A_{2i+1}^{00}-A_{2i-1}^{10}A_{2i+1}^{01}-A_{2i-1}^{01}A_{2i+1}^{10})
+L_0A_{2i-1}^{00}A_{2i+1}^{00}
\}P_{SB-SAT}.
\label{sosbsat}
\end{eqnarray}
\end{widetext}
In above, we have introduced the following notation for the coefficients:
\begin{eqnarray}
L_0=-\frac{(J'_{xx})^2}{2J}\frac{4J+J'}{4J+3J'}
\quad \mbox{and} \quad
L_1=-\frac{(J'_{xx})^2}{4J},
\end{eqnarray}
which enable to write the second-order contribution (\ref{sosbsat}) to the effective Hamiltonian in a more compact form. Next, let us proceed to a notion of the quantum lattice gas achieved through the following transformation: $A_i^{00}=n_i$, $A_i^{11}=(1-n_i)$, $A_i^{01}=a_i^+$, $A_i^{10}=a_i$ ($a_i$ and $a_i^+$ are Pauli operators which anticommute on one site and commute on different sites). The overall effective Hamiltonian can be subsequently rewritten in the particle representation as:
\begin{eqnarray}
H_{eff}&=&\sum_{i=1}^{N/2}P_{SB-SAT}\left\{
(h-h_{c3})n_{2i} 
\right.\nonumber\\&&\left.
{+} (h{-}h_{c3}{+}2L_1)n_{2i+1} 
{+} (L_0{-}2L_1)n_{2i-1}n_{2i+1} 
\right.\nonumber\\&&\left.
{-} L_1(a^+_{2i-1} a_{2i+1} {+} a_{2i-1} a^+_{2i+1})
\right\}P_{SB-SAT}.
\label{eff_SB-SAT}
\end{eqnarray}
The effective quantum lattice-gas model (\ref{eff_SB-SAT}) contains two types of particles. The particles on odd sites are mobile and they hop in between nearest-neighbor odd sites, while the particles on even sites are localized. The projection operator (\ref{b_SB-SAT}) additionally leads to the hard-core repulsion, which blocks the occupation of nearest-neighbor sites. To get the ground state, one has to find such a configuration of the localized particles on the even sites given by the set of occupation numbers $\{n_{2i}\}$, which corresponds to the lowest-energy eigenstate of the quantum subsystem on the odd sites. It is worthy to note that the corresponding quantum part is split into two open chains at each even site $n_{2i}=1$ occupied by the particle, whereas occupation of the neighboring odd sites $2i-1$ and $2i+1$ should be then excluded ($n_{2i-1}=n_{2i+1}=0$). In the following we will show that the energy of the system increases whenever the empty even site changes to the filled one (i.e. $n_{2i}$ changes from 0 to 1). This fact should be proven separately for two cases: $h>h_{c3}$ and $h<h_{c3}$. Let us denote by $E_{0(1)}$ the energy for the empty (filled) even site $n_{2i}=0(1)$. It is quite clear from previous arguments that $E_1=E_L+E_R+(h-h_{c3})$, where $E_L$ and $E_R$ are the lowest energies of the left and right parts of the system split by $n_{2i}=1$ (see Fig.~\ref{fig_eff-scheme}).
The following inequality can be also obtained $E_0\leq E_L + E_R$, which furnishes the proof for $h>h_{c3}$:
\begin{eqnarray}
E_0\leq E_1 + h_{c3} -h < E_1.
\end{eqnarray}
In the opposite case $h<h_{c3}$ we have to use the property that a sum of the ground-state energies of two separate chains and clusters is larger than the ground-state energy of the whole system, which is obtained by joining the separate subsystems together. This property implies a validity of the following inequality
\begin{eqnarray}
E_0 \leq E_L + E_M +  E_R.
\end{eqnarray}
After some algebra one can also show that the following inequality holds for $h<h_{c3}$
\begin{eqnarray}
E_0\leq E_1 + h - h_{c3} + 2L_1 < E_1.
\end{eqnarray}
Accordingly, the ground state should correspond to the particular case with all empty even sites ($n_{2i}=0$ for all $i$), whereas the effective Hamiltonian (\ref{eff_SB-SAT}) of the quantum lattice-gas model then reduces to
\begin{eqnarray}
&&H_{eff}{=}\sum_{i=1}^{N/2}\left\{
 (h{-}h_{c3}{+}2L_1)n_{2i+1} 
{+} (L_0{-}2L_1)n_{2i-1}n_{2i+1}
\right.\nonumber\\&&\left. 
{-} L_1(a^+_{2i-1} a_{2i+1} {+} a_{2i-1} a^+_{2i+1})
\right\}.
\label{hefsbsat}
\end{eqnarray}
The SAT ground state corresponds to the empty state in the particle language ($n_{2i-1}=n_{2i}=0$ for all $i$), while the SB ground state pertinent to the 1/2-plateau emerges when all odd sites are filled by particles and all even sites are being empty ($n_{2i-1}=1$, $n_{2i}=0$ for all $i$). To get the respective values of the critical fields, it is more convenient to convert the effective quantum lattice-gas model (\ref{hefsbsat}) into a pseudospin language. As a matter of fact, one gets the effective Hamiltonian of the spin-1/2 $XXZ$ Heisenberg chain using $\tilde{s}_i^x=(a_{2i-1}^+ + a_{2i-1})/2$, $\tilde{s}_i^y=-i(a_{2i-1}^+ - a_{2i-1})/2$, $\tilde{s}_i^z=a_{2i-1}^+ a_{2i-1}-1/2$:
\begin{eqnarray}
H_{eff}^{xxz}&=&\sum_{i=1}^{N/2}\left(
J^z_{eff}\tilde{s}_i^z \tilde{s}_{i+1}^z
+J^{xx}_{eff} (\tilde{s}_i^x \tilde{s}_{i+1}^x + \tilde{s}_i^y \tilde{s}_{i+1}^y) 
\right.\nonumber\\&&\left.
-h_{eff} \tilde{s}_i^z
\right), 
\nonumber\\
J^z_{eff}&=&\frac{(J'_{xx})^2J'}{J(4J+3J')}>0,
\nonumber\\
J^{xx}_{eff}&=&\frac{(J'_{xx})^2}{2J}>0,
\nonumber\\
h_{eff}&=&h_{c3}-L_0-h.
\label{hefxxz}
\end{eqnarray}
The critical fields for the quantum antiferromagnetic $XXZ$ Heisenberg chain are exactly known: $h_{upper/lower}=\pm(J^z_{eff}+J^{xx}_{eff})=\pm(L_0-4L_1)$. 
Bearing this in mind, the saturation field $h_{sat}$ and the upper critical field $h_{1/2}$ for the 1/2-plateau can be found from the relations
\begin{eqnarray}
&&h_{1/2}=h_{c3}-L_0-h_{upper}=J+J'-\frac{2(J'_{xx})^2J'}{J(4J+3J')},
\nonumber\\
&&h_{sat}=h_{c3}-L_0-h_{lower}=J+J'+\frac{(J'_{xx})^2}{J}.
\end{eqnarray}
It can be seen from Fig.~\ref{fig_phasediag}(b) that both critical fields $h_{1/2}$ and $h_{sat}$ obtained from the perturbative strong-coupling approach quantitatively agree with the numerical DMRG data up to a relative strength between the inter- and intra-dimer couplings $J'/J \approx 0.5$, while the critical field $h_{1/2}$ ($h_{sat}$) is slightly underestimated (overestimated) for greater values of the interaction ratio $J'/J \gtrsim 0.5$. Most importantly, the perturbative expansion around SB-SAT phase boundary predicts the gapless Tomonaga-Luttinger spin-liquid (SL) ground state in a relatively wide range of the magnetic fields $h \in (h_{1/2},h_{sat})$ in spite of the fact that the simplified Ising-Heisenberg model does not exhibit this ground state at all [c.f. Fig.~\ref{fig_phasediag}(a) and (b)]. It should be also pointed out that the spin-1/2 Heisenberg orthogonal-dimer chain undergoes true continuous (second-order) quantum phase transitions at the critical fields $h_{1/2}$ and $h_{sat}$ delimiting a stability region of the SL ground state in contrast with discontinuous (first-order) phase transitions associated with the magnetization jumps between the other fractional plateaux (see Fig. \ref{fig_mag}). Last but not least, the perturbative strong-coupling approach brings a deeper insight into the character of the SL phase, because the number of the odd filled sites within the effective quantum lattice-gas model continuously decreases with increasing of the magnetic field by keeping all even sites empty. When returning back to the spin language this result is taken to mean that the total number of (mobile) singlet states on the horizontal dimers gradually decreases within the SL ground state from its maximum value at the critical field $h_{1/2}$ down to zero at $h_{sat}$ 
while keeping all vertical dimers in the polarized triplet state.  

\section{Conclusions}
\label{sec-conclusions}
The present work dealt with the perturbative strong-coupling calculation for the quantum spin-1/2 Heisenberg orthogonal-dimer chain in a magnetic field, which has been developed from the exactly solved spin-1/2 Ising-Heisenberg orthogonal-dimer chain with the Heisenberg intradimer and Ising interdimer interactions up to the second order. Notably, the quantum spin-1/2 Heisenberg orthogonal-dimer chain represents a paradigmatic example of quantum spin chain with plethora of outstanding quantum ground states, which are manifested in a zero-temperature magnetization curve either as extensive zero, one-quarter and one-half magnetization plateaux, an infinite sequence of tiny fractional $n/(2n+2)$ ($n>1$) magnetization plateaux or the Tomonaga-Luttinger spin-liquid phase. Despite of this complexity, we have convincingly evidenced an impressive numerical accuracy of the strong-coupling approach stemming from the exactly solved Ising-Heisenberg model through a direct comparison of the derived results with the state-of-the-art numerical data obtained within DMRG and ED methods. It has been found that the strong-coupling approach not only substantially improves phase boundaries between the already existing ground states of the idealized Ising-Heisenberg orthogonal-dimer chain, but it also gives rise to completely novel quantum ground states such as the fractional one-third plateau or the Tomonaga-Luttinger spin-liquid phase. 
Based on the effective lattice-gas model at MFI-SB boundary, we presumed that higher-order perturbation terms result in the repulsion interactions of a longer range. It is an indication that other tiny fractional plateaux in between the one-quarter and one-half of the saturation magnetization could be recovered within the higher-order perturbation theory.  

It is also worth noticing that the perturbative strong-coupling approach could be alternatively developed from the limit of isolated dimers as it is shown in Appendix \ref{app-uncor-dimers}. However, this simpler version of the perturbative treatment has serious deficiency in that it does not reproduce in the second order neither one-quarter nor one-third magnetization plateaux. It could be thus concluded that the perturbative strong-coupling method developed from the exactly solved Ising-Heisenberg orthogonal-dimer chain is quite superior with respect to its simplified version derived from the limit of isolated dimers. It therefore appears worthwhile to remark that there exist several exact solutions for the hybrid Ising-Heisenberg models, which could be used as useful starting ground for the perturbative analysis (see Ref.~\onlinecite{pla10} and references cited therein). Quite recently, the similar perturbation procedure starting from the exactly solved spin-1/2 Ising-Heisenberg diamond chain has been applied to corroborate an existence of the Tomonaga-Luttinger spin-liquid phase in between the intermediate one-third plateau and saturation magnetization of the quantum spin-1/2 Heisenberg diamond chain \cite{oleg}. Our further goal is to apply the developed strong-coupling approach to the quantum spin-1/2 Heisenberg model on the Shastry-Sutherland lattice to verify or disprove a presence of the questioned fractional magnetization plateaux by making use of the exact solution reported for the spin-1/2 Ising-Heisenberg model on the Shastry-Sutherland lattice \cite{verk14}.

\begin{acknowledgments}
T.V. acknowledges the financial support provided by the National Scholarship Programme of the Slovak Republic for the Support of Mobility of Students, PhD Students, University Teachers, Researchers and Artists. J.S. acknowledges financial support provided by the grant of The Ministry of Education, Science, Research and Sport of the Slovak Republic under the contract Nos. VEGA 1/0331/15 and VEGA 1/0043/16, as well as, by grants of the Slovak Research and Development Agency provided under Contract Nos. APVV-0097-12 and APVV-14-0073.
\end{acknowledgments}

\appendix
\section{Ground state of the spin-1/2 Heisenberg orthogonal-dimer chain: numerical study}
\label{app-num}
To study the magnetization process and the ground-state phase diagram, we have to distinguish two cases: 
$m\leq m_{sat}/2$ and $m\geq m_{sat}/2$. For the first case Schulenburg and Richter showed that the magnetization curve has a step-like form with the magnetization plateaux at $m/m_{sat}=n/(2n+2)$ for integer $n\geq 0$ \cite{schulenburg02}. They noticed that the singlet state on the vertical dimer defragments the model onto non-interacting parts. The ground-state energy of the phase with the magnetization $m/m_{sat}=n/(2n+2)$ can be found by calculating the energy of the cluster with $n$-subsequent vertical dimers in ${\mathbf S}_{2i}=1$ state \cite{schulenburg02}. We find this energy using the exact diagonalization method of ALPS package \cite{alps} and build the magnetization curve for $m\leq m_{sat}/2$ in Figs. \ref{fig_phasediag},\ref{fig_mag}.

For $m\geq m_{sat}/2$ the ground state corresponds to the situation when all vertical dimers are in the triplet state ${\mathbf S}_{2i}=1$. Using the spin-1 representation for the total spin on vertical dimers \cite{strecka14}, the ground state of the orthogonal-dimer chain can be found among the lowest-energy states of the mixed $\frac{1}{2}-\frac{1}{2}-1$ Heisenberg chain:
\begin{eqnarray}
&&H=\sum_{i=1}^{N/2}\left[
J'( ({\mathbf s}_{2,2i-1}\cdot{\mathbf S}_{2i})+ ({\mathbf S}_{2i}\cdot{\mathbf s}_{1,2i+1}) )
\right.\nonumber\\&&\left.
{+}J({\mathbf s}_{1,2i-1}{\cdot}{\mathbf s}_{2,2i-1})
{-}h(s^z_{1,2i-1}{+}s^z_{2,2i-1}{+}S^z_{2i})
\right],
\end{eqnarray}
where ${\mathbf S}_{2i}$ represents the composite spin-1 particle. To this end, we have performed the DMRG computation using ALPS package \cite{alps} for the systems of $N=128$. 

It should be noted that an analogous numerical approach was used for the orthogonal-dimer chain with triangular clusters \cite{ohanyan12}.

\section{Projection operators}
\label{app-projection}
The correspondence between the spin and projection operators can be found by a straightforward calculation \cite{parkinson79}:
\begin{eqnarray}
s_{1,i}^+&=&\frac{1}{\sqrt2}(A_i^{12}+A_i^{23}-A_i^{10}+A_i^{03}),
\nonumber\\
s_{2,i}^+&=&\frac{1}{\sqrt2}(A_i^{12}+A_i^{23}+A_i^{10}-A_i^{03}),
\nonumber\\
s_{1,i}^-&=&\frac{1}{\sqrt2}(A_i^{21}+A_i^{32}-A_i^{01}+A_i^{30}),
\nonumber\\
s_{2,i}^-&=&\frac{1}{\sqrt2}(A_i^{21}+A_i^{32}+A_i^{01}-A_i^{30}),
\nonumber\\
s_{1,i}^z&=&\frac{1}{2}(A_i^{11}-A_i^{33}+A_i^{02}+A_i^{20}),
\nonumber\\
s_{2,i}^z&=&\frac{1}{2}(A_i^{11}-A_i^{33}-A_i^{02}-A_i^{20}).
\end{eqnarray}

\section{Coefficients of unitary transformation}
\label{app-UT}
Coefficients of the unitary transformation
$\cos\frac{\alpha_{2i+1}}{2}$, $\sin\frac{\alpha_{2i+1}}{2}$ can be obtained from Eq.(\ref{u-transform}) using the formulae for 
trigonometric functions of half argument:
\begin{eqnarray}
\cos\frac{\alpha_{2i+1}}{2}&=&1 + (c_1^+ -1)\delta(|S_{2i}^z - S_{2i+2}^z|-1) 
\nonumber\\
&&+ (c_2^+ -1)\delta(|S_{2i}^z - S_{2i+2}^z|-2),
\nonumber\\
\sin\frac{\alpha_{2i+1}}{2}&=&c_1^-\delta'(|S_{2i}^z - S_{2i+2}^z|-1) 
\nonumber\\&&
+ c_2^-\delta'(|S_{2i}^z - S_{2i+2}^z|-2),
\nonumber\\
c_1^{\pm}&=&\frac{1}{\sqrt2}\sqrt{1\pm\frac{J}{\sqrt{J^2+J'^2}} },
\nonumber\\
c_2^{\pm}&=&\frac{1}{\sqrt2}\sqrt{1\pm\frac{J}{\sqrt{J^2+4J'^2}} },
\end{eqnarray}
\begin{eqnarray}
\delta(|S_{2i}^z - S_{2i+2}^z|-1)
&=&((S_{2i}^z)^2 - (S_{2i+2}^z)^2)^2
\nonumber\\
&{=}&(A_{2i}^{11}{+}A_{2i}^{33})(A_{2i+2}^{00}{+}A_{2i+2}^{22}) 
\nonumber\\&&
{+} (A_{2i}^{00}{+}A_{2i}^{22})(A_{2i+2}^{11}{+}A_{2i+2}^{33}),
\nonumber\\
\delta(|S_{2i}^z - S_{2i+2}^z|-2)
&=&\frac{1}{2}S_{2i}^z S_{2i+2}^z(S_{2i}^z S_{2i+2}^z -1)
\nonumber\\
&=&A_{2i}^{11}A_{2i+2}^{33} + A_{2i}^{33}A_{2i+2}^{11},
\nonumber\\
\delta'(|S_{2i}^z {-} S_{2i+2}^z|-1)&{=}&(S_{2i}^z {-} S_{2i+2}^z)\delta(|S_{2i}^z {-} S_{2i+2}^z|{-}1)
\nonumber\\
&{=}&(A_{2i}^{11}{-}A_{2i}^{33})(A_{2i+2}^{00}{+}A_{2i+2}^{22}) 
\nonumber\\
&&- (A_{2i}^{00}+A_{2i}^{22})(A_{2i+2}^{11}-A_{2i+2}^{33}),
\nonumber\\
\delta'(|S_{2i}^z {-} S_{2i+2}^z|{-}2)&=&\frac{1}{2}(S_{2i}^z {-} S_{2i+2}^z)\delta(|S_{2i}^z {-} S_{2i+2}^z|{-}2)
\nonumber\\&&
=A_{2i}^{11}A_{2i+2}^{33} - A_{2i}^{33}A_{2i+2}^{11}.
\label{kron-delta}
\end{eqnarray}

\begin{widetext}
\section{Effective Hamiltonian at SD-MFI boundary}
\label{app-SD-MFI}
For obtaining the second order perturbation defined by Eq. (\ref{eq_pert}) we have to calculate at first:
\begin{eqnarray}
\tilde{H}_{2i}^{(1)}P_{SD-MFI}{=}\frac{J'_{xx}}{2}(c_1^+ {+} c_1^-) 
[A_{2i-1}^{10}A_{2i}^{21}(c_1^+A_{2i+1}^{00} {-} c_1^-A_{2i+1}^{20}) 
{-} (c_1^+A_{2i-1}^{00} {+} c_1^-A_{2i-1}^{20})A_{2i}^{21}A_{2i+1}^{10} ] P_{SD-MFI},
\nonumber 
\end{eqnarray}
\begin{eqnarray}
R_s\tilde{H}_{2i}^{(1)}P_{SD-MFI}&{=}&\frac{J'_{xx}}{2}(c_1^+ {+} c_1^-)
\left\{ \frac{c_1^+ A_{2i-1}^{11}A_{2i}^{22}A_{2i+1}^{00}}{\Delta E_{2l}(120)}A_{2i-1}^{10}A_{2i}^{21}A_{2i+1}^{00}
{-} \frac{c_1^- A_{2i-1}^{11}A_{2i}^{22}A_{2i+1}^{22}}{\Delta E_{2l}(122)}A_{2i-1}^{10}A_{2i}^{21}A_{2i+1}^{20}
\right.
\nonumber\\
&&\left.
{-} \frac{c_1^+ A_{2i-1}^{00}A_{2i}^{22}A_{2i+1}^{11}}{\Delta E_{2l}(021)}A_{2i-1}^{00}A_{2i}^{21}A_{2i+1}^{10}
{-} \frac{c_1^- A_{2i-1}^{22}A_{2i}^{22}A_{2i+1}^{11}}{\Delta E_{2l}(221)}A_{2i-1}^{20}A_{2i}^{21}A_{2i+1}^{10}
\right\}P_{SD{-}MFI},
\nonumber\\
\Delta  E_{2l}(120) &=& -\sqrt{J^2+J'^2}, \;
\Delta  E_{2l}(122) = -J-\sqrt{J^2+J'^2}.
\end{eqnarray}
Here $\Delta  E_{2l}(n_{2l-1},n_{2l},n_{2l+1})=E_{GS}- E_{2l}(n_{2l-1},n_{2l},n_{2l+1})$ is the difference between the ground-state energy and the energy of the excitation on three consequtive coupled dimers $2l-1$, $2l$, $2l+1$ in states $n_{2l-1}$, $n_{2l}$, $n_{2l+1}$. 
Using the relation $P_{SD-MFI}\tilde{H}_{2i}^{(1)}=(\tilde{H}_{2i}^{(1)}P_{SD-MFI})^+$, we can obtain the second-order term in (\ref{eq_pert}) explicitly:
\begin{eqnarray}
&&H_{eff}^{(2)}= \sum_{i=1}^{N/2}P_{SD-MFI}\tilde{H}_{2i}^{(1)} R_s \tilde{H}_{2i}^{(1)}P_{SD-MFI}
=\sum_{i=1}^{N/2}P_{SD-MFI} h^{(2)} A_{2i}^{11}P_{SD-MFI},
\nonumber\\
&&h^{(2)}=\frac{(J'_{xx})^2(1+2c_1^+c_1^-)}{2}\left\{
\frac{(c_1^+)^2}{\Delta E_{2l}(120)} + \frac{(c_1^-)^2}{\Delta E_{2l}(122)}
\right\}.
\end{eqnarray}

\section{Effective Hamiltonian at MFI-SB boundary}
\label{app-MFI-SB}
Following the same procedure as in Appendix \ref{app-SD-MFI}, we get:
\begin{eqnarray}
\tilde{H}_{2i}^{(1)}P_{MFI-SB}&=&\frac{J'_{xx}}{2}
\left\{
(A_{2i-2}^{11}+(c_1^+ + c_1^-)A_{2i-2}^{00} ) A_{2i-1}^{10} A_{2i}^{21} (c_1^+ A_{2i+1}^{00} - c_1^-A_{2i+1}^{20} )
\right.
\nonumber\\
&&\left.
+(c_1^+ A_{2i-1}^{00} + c_1^-A_{2i-1}^{20} ) A_{2i}^{21} A_{2i+1}^{10} (A_{2i+2}^{11}-(c_1^+ + c_1^-)A_{2i+2}^{00} )
\right\}P_{MFI-SB},
\label{h-p-mfi}
\end{eqnarray}
\begin{eqnarray}
R_s\tilde{H}_{2i}^{(1)}P_{MFI-SB}&=&\frac{J'_{xx}}{2}
\left\{
\left[
c_1^+\left(
\frac{ A_{2i-2}^{11} A_{2i-1}^{11} A_{2i}^{22} A_{2i+1}^{00} A_{2i+2}^{00} }{\Delta E_{2i}(11200)}
+\frac{ A_{2i-2}^{11} A_{2i-1}^{11} A_{2i}^{22} A_{2i+1}^{00} A_{2i+2}^{11} }{\Delta E_{2i}(11201)}\right)
\right.\right.
\nonumber\\
&&\left.\left.
-c_1^-\left(
\frac{ A_{2i-2}^{11} A_{2i-1}^{11} A_{2i}^{22} A_{2i+1}^{22} A_{2i+2}^{00} }{\Delta E_{2i}(11220)}
+\frac{ A_{2i-2}^{11} A_{2i-1}^{11} A_{2i}^{22} A_{2i+1}^{22} A_{2i+2}^{11} }{\Delta E_{2i}(11221)}\right) A_{2i+1}^{20}
\right.\right.
\nonumber\\
&&\left.\left.
+c_1^+(c_1^+ + c_1^-)\left(
\frac{ A_{2i-2}^{00} A_{2i-1}^{11} A_{2i}^{22} A_{2i+1}^{00} A_{2i+2}^{00} }{\Delta E_{2i}(01200)}
+\frac{ A_{2i-2}^{00} A_{2i-1}^{11} A_{2i}^{22} A_{2i+1}^{00} A_{2i+2}^{11} }{\Delta E_{2i}(01201)}\right)
\right.\right.
\nonumber\\
&&\left.\left.
-c_1^- (c_1^+ + c_1^-)\left(
\frac{ A_{2i-2}^{00} A_{2i-1}^{11} A_{2i}^{22} A_{2i+1}^{22} A_{2i+2}^{00} }{\Delta E_{2i}(01220)}
+\frac{ A_{2i-2}^{00} A_{2i-1}^{11} A_{2i}^{22} A_{2i+1}^{22} A_{2i+2}^{11} }{\Delta E_{2i}(01221)}\right)A_{2i+1}^{20}
\right]A_{2i-1}^{10} A_{2i}^{21}
\right.
\nonumber\\
&&\left.
+\left[
c_1^+\left(
\frac{ A_{2i-2}^{00} A_{2i-1}^{00} A_{2i}^{22} A_{2i+1}^{11} A_{2i+2}^{11} }{\Delta E_{2i}(00211)}
+\frac{ A_{2i-2}^{11} A_{2i-1}^{00} A_{2i}^{22} A_{2i+1}^{11} A_{2i+2}^{11} }{\Delta E_{2i}(10211)}\right)
\right.\right.
\nonumber\\
&&\left.\left.
+c_1^-\left(
\frac{ A_{2i-2}^{00} A_{2i-1}^{22} A_{2i}^{22} A_{2i+1}^{11} A_{2i+2}^{11} }{\Delta E_{2i}(02211)}
-\frac{ A_{2i-2}^{11} A_{2i-1}^{22} A_{2i}^{22} A_{2i+1}^{11} A_{2i+2}^{11} }{\Delta E_{2i}(12211)}\right) A_{2i-1}^{20}
\right.\right.
\nonumber\\
&&\left.\left.
-c_1^+(c_1^+ + c_1^-)\left(
\frac{ A_{2i-2}^{00} A_{2i-1}^{00} A_{2i}^{22} A_{2i+1}^{11} A_{2i+2}^{00} }{\Delta E_{2i}(00210)}
+\frac{ A_{2i-2}^{11} A_{2i-1}^{00} A_{2i}^{22} A_{2i+1}^{11} A_{2i+2}^{00} }{\Delta E_{2i}(10210)}\right)
\right.\right.
\nonumber\\
&&\left.\left.
-c_1^- (c_1^+ + c_1^-)\left(
\frac{ A_{2i-2}^{00} A_{2i-1}^{22} A_{2i}^{22} A_{2i+1}^{11} A_{2i+2}^{00} }{\Delta E_{2i}(02210)}
+\frac{ A_{2i-2}^{11} A_{2i-1}^{22} A_{2i}^{22} A_{2i+1}^{11} A_{2i+2}^{00} }{\Delta E_{2i}(12210)}\right)A_{2i-1}^{20}
\right]A_{2i}^{21} A_{2i+1}^{10}
\right\}
\nonumber\\
&&\times P_{MFI-SB},
\label{Rs-h-p-mfi}
\end{eqnarray}
\begin{eqnarray}
\Delta E_{2i}(01200)&=&-\sqrt{J^2+J'^2}, \;
\Delta E_{2i}(01220)=\Delta E_{2i}(01221)=-J-\sqrt{J^2+J'^2},
\nonumber\\
\Delta E_{2i}(11201)&=&\frac{1}{2}(-3J-J'+\sqrt{J^2+J'^2}), \;
\Delta E_{2i}(11221)=\Delta E_{2i}(11220)=\frac{1}{2}(-3J-J'-\sqrt{J^2+J'^2}),
\nonumber\\
\Delta E_{2i}(11200)&=&\frac{1}{2}(-J-J'-\sqrt{J^2+J'^2}), \;
\Delta E_{2i}(01201)=-J.
\end{eqnarray}
Here we introduce $\Delta E_{2i}(n_{2i-2},n_{2i-1},n_{2i},n_{2i+1},n_{2i+2})$ analogously to the notation in Appendix~\ref{app-SD-MFI}.
Combining Eqs. (\ref{h-p-mfi}) and (\ref{Rs-h-p-mfi}), we get for the second order term the following result:
\begin{eqnarray}
H_{eff}^{(2)}&=& P_{MFI-SB}\sum_{i=1}^{N/2}\tilde{H}_{2i}^{(1)} R_s \sum_{j=1}^{N/2}\tilde{H}_{2j}^{(1)}P_{MFI-SB}
\nonumber\\
&=& \frac{(J'_{xx})^2}{4}\sum_{i=1}^{N/2} P_{MFI-SB}
\left\{
2(c_1^+ + c_1^-)^2
\left(\frac{(c_1^+)^2}{\Delta E_{2i}(01200)} + \frac{(c_1^-)^2}{\Delta E_{2i}(01220)}
\right)A_{2i-2}^{00}A_{2i+2}^{00}
\right.
\nonumber\\
&&\left.
+2\left(\frac{(c_1^+)^2}{\Delta E_{2i}(11201)} + \frac{(c_1^-)^2}{\Delta E_{2i}(11221)}
\right)A_{2i-2}^{11}A_{2i+2}^{11}
+\left(
\frac{(c_1^+)^2}{\Delta E_{2i}(11200)} + \frac{(c_1^-)^2}{\Delta E_{2i}(11220)}
\right.\right.
\nonumber\\
&&+\left.\left.
(c_1^+ + c_1^-)^2
\left(\frac{(c_1^+)^2}{\Delta E_{2i}(01201)} + \frac{(c_1^-)^2}{\Delta E_{2i}(01221)}
\right)
\right)
(A_{2i-2}^{11}A_{2i+2}^{00}+A_{2i-2}^{00}A_{2i+2}^{11})
\right\}A_{2i}^{11} P_{MFI-SB}.
\end{eqnarray}

\section{Effective Hamiltonian at SB-SAT boundary}
\label{app-SB-SAT}
Analogously the following expressions for the case of the SB-SAT boundary can be obtained:
\begin{eqnarray}
\tilde{H}_{2i}^{(1)}P_{SB-SAT}&=&\frac{J'_{xx}}{2}\left[
A_{2i-2}^{11}A_{2i-1}^{10}A_{2i}^{21}A_{2i+1}^{11} {-} A_{2i-1}^{11}A_{2i}^{21}A_{2i+1}^{10}A_{2i+2}^{11}
{+}A_{2i-2}^{11}A_{2i-1}^{10}A_{2i}^{21}(c_1^+ A_{2i+1}^{00} {-} c_1^-A_{2i+1}^{20})A_{2i+2}^{11}
\right.
\nonumber\\ 
&&
\left.
-A_{2i-2}^{11}(c_1^+ A_{2i-1}^{00} + c_1^-A_{2i-1}^{20})A_{2i}^{21}A_{2i+1}^{10}A_{2i+2}^{11}
\right] P_{SB-SAT},
\\
R_s\tilde{H}_{2i}^{(1)}P_{SB-SAT}&=&\frac{J'_{xx}}{2}
\left\{\left[
\frac{ A_{2i-2}^{11} A_{2i-1}^{11} A_{2i}^{22} A_{2i+1}^{11} A_{2i+2}^{00} }{\Delta E_{2i}(11210)}
+\frac{ A_{2i-2}^{11} A_{2i-1}^{11} A_{2i}^{22} A_{2i+1}^{11} A_{2i+2}^{11} }{\Delta E_{2i}(11211)}
\right.\right.
\nonumber\\
&&\left.\left.
+\frac{ c_1^+ A_{2i-2}^{11} A_{2i-1}^{11} A_{2i}^{22} A_{2i+1}^{00} A_{2i+2}^{11} }{\Delta E_{2i}(11201)}
-\frac{ c_1^- A_{2i-2}^{11} A_{2i-1}^{11} A_{2i}^{22} A_{2i+1}^{20} A_{2i+2}^{11} }{\Delta E_{2i}(11221)}
\right]A_{2i-1}^{10} A_{2i}^{21}
\right.
\nonumber\\
&&\left.
-\left[
\frac{ A_{2i-2}^{00} A_{2i-1}^{11} A_{2i}^{22} A_{2i+1}^{11} A_{2i+2}^{11} }{\Delta E_{2i}(01211)}
+\frac{ A_{2i-2}^{11} A_{2i-1}^{11} A_{2i}^{22} A_{2i+1}^{11} A_{2i+2}^{11} }{\Delta E_{2i}(11211)}
\right.\right.
\nonumber\\
&&\left.\left.
{+}\frac{ c_1^+ A_{2i-2}^{11} A_{2i-1}^{00} A_{2i}^{22} A_{2i+1}^{11} A_{2i+2}^{11} }{\Delta E_{2i}(10211)}
{+}\frac{ c_1^- A_{2i-2}^{11} A_{2i-1}^{20} A_{2i}^{22} A_{2i+1}^{11} A_{2i+2}^{11} }{\Delta E_{2i}(12211)}
\right]A_{2i}^{21} A_{2i+1}^{10}
\right\}P_{SB{-}SAT},
\\
\Delta E_{2i}(11211)&=&\Delta E_{2i}(11210)=-J,
\nonumber\\
\Delta E_{2i}(11201)&=&-\frac{1}{2}(3J+J'-\sqrt{J^2+J'^2}),
\nonumber\\
\Delta E_{2i}(11221)&=&-\frac{1}{2}(3J+J'+\sqrt{J^2+J'^2}).
\end{eqnarray}
Then, the second-order perturbation term is as follows
\begin{eqnarray}
H_{eff}^{(2)}&=&\frac{(J'_{xx})^2}{4}\sum_{i=1}^{N/2}P_{SB-SAT}\left\{
-\frac{(A_{2i-1}^{10}A_{2i+1}^{01} + A_{2i-1}^{01}A_{2i+1}^{10})}{\Delta E_{2i}(11211)}
+\frac{(A_{2i-1}^{00}A_{2i+1}^{11}+A_{2i-1}^{11}A_{2i+1}^{00})}{\Delta E_{2i}(11211)}
\right.\nonumber\\&&\left.
+\left(\frac{2(c_1^+)^2}{\Delta E_{2i}(11201)}+\frac{2(c_1^-)^2}{\Delta E_{2i}(11221)}
\right)A_{2i-1}^{00}A_{2i+1}^{00}
\right\}P_{SB-SAT}.
\end{eqnarray}
We introduce the further notations:
\begin{eqnarray}
L_0&=&-\frac{(J'_{xx})^2}{2}\left(
\frac{(c_1^+)^2}{\Delta E_{2i}(11201)} + \frac{(c_1^-)^2}{\Delta E_{2i}(11221)}
\right)=-\frac{(J'_{xx})^2}{2J}\frac{4J+J'}{4J+3J'},
\nonumber\\
L_1&=&-\frac{(J'_{xx})^2}{4\Delta E_{2i}(11211)}=-\frac{(J'_{xx})^2}{4J},
\end{eqnarray}
then
\begin{eqnarray}
H_{eff}^{(2)}=\sum_{i=1}^{N/2}P_{SB-SAT}\{
L_1(A_{2i-1}^{00}A_{2i+1}^{11}+A_{2i-1}^{11}A_{2i+1}^{00}-A_{2i-1}^{10}A_{2i+1}^{01}-A_{2i-1}^{01}A_{2i+1}^{10})
+L_0A_{2i-1}^{00}A_{2i+1}^{00}
\}P_{SB-SAT}.
\end{eqnarray}
\end{widetext}

\section{Strong-coupling approach from the limit of isolated dimers}
\label{app-uncor-dimers}
Here we consider the strong coupling expansion around the limit of noninteracting dimers. In this case, we  treat the total interdimer coupling as a perturbation:
\begin{eqnarray}
\label{gen_ham2}
H&{=}& H^{(0)} + H^{(1)} = \sum_{i=1}^N H^{(0)}_i +  \sum_{i=1}^N H^{(1)}_i,
\nonumber\\
H^{(0)}_{i}&{=}&
J({\mathbf s}_{1,i}\cdot{\mathbf s}_{2,i}) - h_c(s_{1,i}^z+s_{2,i}^z),
\nonumber\\
H^{(1)}_{2i}&{=}& (h_c - h)(s_{1,2i}^z+s_{2,2i}^z)
\nonumber\\&&
+J'({\mathbf s}_{2,2i-1} + {\mathbf s}_{1,2i+1})\cdot({\mathbf s}_{1,2i} + {\mathbf s}_{2,2i}),
\nonumber\\
H^{(1)}_{2i+1}&{=}& (h_c - h)(s_{1,2i+1}^z+s_{2,2i+1}^z),
\end{eqnarray}
The Hamiltonian of isolated dimers $H^{(0)}$ exhibits two phases in the ground state. If $h<h_c=J$ the ground state consists of singlet dimers, otherwise it is polarized in the direction of the field. At the boundary the ground state is macroscopically degenerate where each dimer can be in any of two states singlet or triplet. 
We rewrite the Hamiltonian in terms of the projection operators (\ref{proj-operators}) and perform the many-body perturbation theory (\ref{eq_pert}) to get the following effective Hamiltonian up to second order terms:
\begin{eqnarray}
H_{eff}&=&\sum_{i=1}^N\left(\frac{J'}{2}+\frac{(J')^2}{4J}\right)A_i^{11} A_{i+1}^{11} 
\nonumber\\&{-}&
\sum_{i=1}^{N/2}\left[(h{-}h_c)A_{2i-1}^{11} {+} \left(h{-}h_c{+}\frac{(J')^2}{2J}\right)A_{2i}^{11}\right]
\nonumber\\
&{+}&\frac{(J')^2}{4J}\sum_{i=1}^{N/2}(A_{2i-1}^{01}A_{2i+1}^{10}{+}A_{2i+1}^{01}A_{2i-1}^{10})A_{2i}^{11}.
\end{eqnarray}
Similarly to the arguments of the previous section, it can be shown that SD phase corresponding to $\prod_{i=1}^N|0\rangle_i$ is the ground state until $h<\tilde{h}(0\to 1/2)$, where 
\begin{eqnarray}
\tilde{h}(0\to 1/2)=J-\frac{(J')^2}{2J}.
\end{eqnarray}
Above $h=\tilde{h}(0\to 1/2)$ the ground state should be sought in the subspace where all vertical dimers are in polarized state 
$\prod_{i=1}^{N/2}|1\rangle_{2i}$. 
Thus, the effective Hamiltonian for odd sites (horizontal dimers) can be presented as a hard-core Bose gas with an infinite on-site repulsion:
\begin{eqnarray}
H^{odd}_{eff}&=&\sum_{i=1}^{N/2}\left[\left(J+J'+\frac{(J')^2}{2J}-h\right)b_i^+ b_i 
\right.\nonumber\\&&
\left.
+ \frac{(J')^2}{4J} (b_i^+ b_{i+1}+ b_{i+1}^+ b_i)\right].
\end{eqnarray}
Here we introduce new creation and annihilation operators $b_i^+=A_{2i-1}^{10}$, $b_i=A_{2i-1}^{01}$ which commute on different sites and anticommute on the same site. $b_i^+=A_{2i-1}^{10}$ turns the singlet state $|{0}\rangle_{2i-1}$ with $S^z_{2i-1}=0$ 
to the triplet state $|{1}\rangle_{2i-1}$ with $S^z_{2i-1}=1$. Thus, it creates the magnon excitation, and $n_i=b_i^+ b_i$ counts the number of magnons. The averaged magnetization of the horizontal dimers can be found from the density of the magnon excitations:
$\langle m_{h}\rangle=(2/N)\sum_{i=1}^{N/2}\langle n_i\rangle$. The model is exactly solved by Jordan-Wigner transformation \cite{lsm} and the averaged averaged magnetization of the horizontal dimers is given as follows:
\begin{widetext}
\begin{eqnarray}
\langle m_h\rangle&=&
\left\{
\begin{array}{ll}
0, & {\rm if}\: h<\tilde{h}_{1/2},\\
\frac{1}{\pi}\arccos\left(1+ 2\frac{J}{J'}+ 2\left(\frac{J}{J'}\right)^2-\frac{2Jh}{J'2} \right), & {\rm if}\: \tilde{h}_{1/2}<h<\tilde{h}_{sat},\\
1, & {\rm if}\: h>\tilde{h}_{sat},
\end{array},
\right.
\\
\tilde{h}_{1/2}&=&J+J', \qquad
\tilde{h}_{sat}=J+J'+\frac{(J')^2}{J}.
\end{eqnarray}
\end{widetext}
As a result within the considered approximation the system is in the SB phase for $h(0\to1/2)<h<\tilde{h}_{1/2}$, in the spin-liquid phase for $\tilde{h}_{1/2}<h<\tilde{h}_{sat}$, and in the saturated phase for $h>\tilde{h}_{sat}$. 
Comparing with the numerical results for the Heisenberg model we see that the fractional plateau at 1/4 as well as a series of tiny plateaux between 1/2 and 1/4 is missing. The lower critical field for the spin-liquid state coincides with the SB-SAT boundary of the Ising-Heisenberg model, whereas the critical field $\tilde{h}(0\to1/2)$ is quite close to the SD-MFI boundary of the Ising-Heisenberg model.

\end{document}